\documentclass[twoside,fleqn]{article}
\usepackage{a4}
\usepackage{graphicx}
\usepackage[final]{epsfig}
\usepackage{amssymb}
\begin{document}
\begin{center}
{\bf \LARGE Differentially rotating disks of dust:\\ 
Arbitrary rotation law\\[1cm]}\end{center}
{\bf Marcus Ansorg\footnote{Friedrich-Schiller-Universit\"{a}t Jena,  
Theoretisch-Physikalisches Institut, Max-Wien-Platz 1, 07743 Jena, Germany, 
E-mail: ansorg@tpi.uni-jena.de}\\[1cm]}
\begin{center}
{\bf Abstract}
\end{center}
\begin{quotation}\noindent
In this paper, solutions to the Ernst equation are investigated that depend on 
two real analytic 
functions defined on the interval $[0,1]$. These solutions are introduced by a 
suitable limiting process of 
B\"{a}cklund transformations applied to seed solutions of the Weyl class.\\ 
It turns out that this class of solutions contains the general relativistic 
gravitational field
of an arbitrary differentially rotating disk of dust, for which a continuous 
transition to some Newtonian disk
exists. It will be shown how for given boundary conditions (i.~e.~proper surface
mass density {\em or} angular velocity of the disk) 
the gravitational field can be approximated in terms of the above solutions. 
Furthermore, particular examples will be discussed, including disks with a 
realistic profile for
the angular velocity and more exotic disks possessing two spatially separated 
ergoregions.     
\end{quotation}
\newpage
\section{Introduction}
Differentially rotating disks of dust have already been studied by Ansorg and 
Meinel \cite{AM00}. 
They considered the class of hyperelliptic solutions to the Ernst equation 
introduced by Meinel and Neugebauer \cite{MN96}, see also 
\cite{Kor89}-\cite{MN97}.
These hyperelliptic solutions depend on a number 
of complex parameters and a real potential function. Ansorg and Meinel 
concentrated on the 
case in which {\em one} complex parameter can be prescribed. They determined 
the real potential function
in order to satisfy a particular boundary condition valid for all disks of 
dust. To generate their solutions, 
they used Neugebauer's and Meinel's rigorous solution \cite{NM93,NM95,NKM96} 
to the boundary value problem of a 
rigidly rotating disk of dust which also belongs to the hyperelliptic class.\\
A subclass of Ansorg's and Meinel's solutions is made up of B\"{a}cklund 
transforms of seed
solutions of the Weyl class\footnote{The construction of solutions to the 
Ernst equation 
by means of B\"{a}cklund transformations belongs to the powerful analytic 
methods 
developed by several authors \cite{Mais78}-\cite{Hauserc}. For a detailed 
introduction see \cite{N96}.}.
Solutions of this type are of particular interest since their 
mathematical structure is much simpler than that of the more general 
hyperelliptic solutions. \\
With this in mind, the  following questions arise: 
\begin{itemize}
\item Is it possible to find solutions corresponding to more general 
differentially rotating disks of dust by 
      increasing the number of prescribed complex parameters?
\item If so, is there a rapidly converging method for approximating 
arbitrary differentially 
rotating disks of dust with given boundary conditions (i.~e.~proper mass 
density {\em or} angular velocity)?
\item Is it perhaps possible to construct such a method by restriction to 
the much simpler solutions of the B\"{a}cklund 
type?
\end{itemize}
To answer these questions, the paper is organized as follows. In the first 
section the metric tensor, Ernst equation, and 
boundary conditions are introduced and the class of solutions of the 
B\"{a}cklund type is represented. As will be discussed in 
the second section, the properties of these solutions can be used to obtain 
more general solutions by a suitable 
limiting process. Since these more general solutions depend on two real 
analytic functions defined on the interval 
$[0,1]$, a rapidly converging numerical scheme to satisfy arbitrary boundary 
conditions for disks of dust can be created. This is depicted 
in the third section. Finally, the fourth section contains particular examples 
of differentially rotating disks of dust, 
including disks with a realistic profile for the angular velocity and more 
exotic disks
possessing two spatially separated ergoregions.\\
In what follows, units are used in which the velocity of light as well as 
Newton's constant of gravitation are 
equal to 1.
\subsection{Metric Tensor, Ernst equation, and boundary conditions}
The metric tensor for axisymmetric stationary and asymptotically flat 
space-times reads as follows 
in Weyl-Papapetrou-coordinates $(\rho,\zeta,\varphi,t)$:
\[ds^2=e^{-2U}[e^{2k}(d\rho^2+d\zeta^2)+\rho^2d\varphi^2]
-e^{2U}(dt+a\;d\varphi)^2\,.\]
For this line element, the vacuum field equations 
are equivalent to a single complex equation -- the so-called Ernst 
equation \cite{Ernst,Ernsta}
   \begin{equation}
   \label{G1.1}
   (\Re f)\; \triangle f=({\bf \nabla} f)^2\,,
   \end{equation}
   \[\triangle=\frac{\partial^2}{\partial\rho^2}+\frac{1}{\rho}
   \frac{\partial}{\partial\rho}+\frac{\partial^2}{\partial\zeta^2},
   \qquad \nabla=\left(\frac{\partial}{\partial\rho},\frac{\partial}
   {\partial\zeta}\right),\]
   where the Ernst potential $f$ is given by
   \begin{equation}
   \label{G1.2}
   f=e^{2U}+\mbox{i}\,b\quad\mbox{with}\quad b_{,\zeta}=\frac{e^{4U}}
   {\rho}\,a_{,\rho}\;,\quad
   b_{,\rho}=-\frac{e^{4U}}{\rho}\,a_{,\zeta}.
   \end{equation}
The remaining function $k$ can be calculated from the Ernst potential $f$ 
by a line integration:
   \[\frac{k_{,\rho}}{\rho}=(U_{,\rho})^2-(U_{,\zeta})^2
   +\frac{1}{4}e^{-4U}[(b_{,\rho})^2-(b_{,\zeta})^2]\]
   \[\frac{k_{,\zeta}}{\rho}=2U_{,\rho} U_{,\zeta}
   +\frac{1}{2}e^{-4U}b_{,\rho} b_{,\zeta}.\]
To obtain the boundary conditions for differentially rotating disks of dust, 
one has to consider
the field equations for an energy-momentum-tensor
\[T^{ik}=\epsilon u^i u^k=\sigma_p(\rho)e^{U-k}\delta(\zeta)u^i u^k,\]
where $\epsilon$ and $\sigma_p$ stand for the energy-density and the invariant 
(proper) 
surface mass-density, respectively, $\delta$ is the usual Dirac 
delta-distribution, and $u^i$ 
denotes the four-velocity of the dust material\footnote{$u^i$ 
has only $\varphi$- and $t$- components.}.  \\
Integration of the corresponding field equations from the 
lower to the upper side of the disk (with coordinate radius $\rho_0$) yields 
 the conditions (see \cite{TKl}, pp. 81-83)
\begin{equation}
   \label{G1.3}
2\pi\sigma_p=e^{U-k}(U_{,\zeta}+\frac{1}{2}Q)
\end{equation}
\begin{equation}
   \label{G1.4}
e^{4U}Q^2+Q(e^{4U})_{,\zeta}+(b_{,\rho})^2=0
\end{equation}
for $\zeta=0^+,\;0\leq\rho\leq\rho_0$ and
\begin{equation}
   \label{G1.5}
Q=-\rho e^{-4U}[b_{,\rho}b_{,\zeta}+(e^{2U})_{,\rho}(e^{2U})_{,\zeta}].
\end{equation}
Note that boundary condition (\ref{G1.4}) for the Ernst potential $f$ does not 
involve the 
surface mass-density $\sigma_p$. This condition comes from the nature of the 
material 
the disk is made of. Therefore, equation (\ref{G1.4}) will be referred to as 
the 
{\it dust-condition}.\\
Instead of prescribing the proper surface mass-density $\sigma_p$ 
[which leads to the boundary condition (\ref{G1.3})] 
one can alternatively assume a given angular velocity 
$\Omega=\Omega(\rho)=u^{\varphi}/u^t$ of the disk which results 
in the boundary condition ($\zeta=0^+$, $0\leq\rho\leq\rho_0$):
\begin{equation}
   \label{G1.6}
\Omega=\frac{Q}{a_{,\zeta}-a\;Q}\,.
\end{equation}
The following requirements due to symmetry conditions and asymptotical 
flatness complete the set of boundary conditions:
\begin{itemize}
\item Regularity at the rotation axis is guaranteed by 
      \[\frac{\partial f}{\partial\rho}(0,\zeta)=0.\] 
\item At infinity asymptotical flatness is realized by $U\to 0$ and $a\to 0$. 
For the 
potential $b$ this has the consequence $b\to b_\infty=\mbox{const}$. Without 
loss of generality,
this constant can be set to $0$, i.e. $f\to 1$ at infinity. 
\item Finally, reflectional symmetry with respect to the plane 
$\zeta=0$ is assumed, i.e. 
$f(\rho,-\zeta)=\overline{f(\rho,\zeta)}$ (with a bar denoting complex 
conjugation).

\end{itemize}      
\subsection{Solutions of the B\"{a}cklund type}
   For a given integer $q\geq 1$, a set 
   $\{Y_1,\ldots,Y_q\}=\{Y_\nu\}_q$\footnote{In the following, the notation 
   $\{Y_1,\ldots,Y_q\}$ will be abbreviated by $\{Y_\nu\}_q$.}
   of complex parameters, and a real analytic function $g$ defined on 
   the interval $[0,1],$ the following expression 
   \begin{eqnarray}
   \label{G1.7}
   f=f_0\,\,\frac{\left|\begin{array}{cccccccc} 
                           1&                1     &                         1& 
                                          1     &                         1  
                                          &\cdots&1&                    1   
                                          \\[2mm]
                          -1&\alpha_1\lambda_1     &\alpha_1^*\lambda_1^*    
                           &\alpha_2\lambda_2     &\alpha_2^*\lambda_2^*     
                             &\cdots&\alpha_q\lambda_q&\alpha_q^*\lambda_q^*  
                              \\[2mm]
                           1&\lambda_1^2           &(\lambda_1^*)^2&
                           \lambda_2^2           &(\lambda_2^*)^2            
                            &\cdots&\lambda_q^2&(\lambda_q^*)^2         
                             \\[2mm]
                          -1&\alpha_1\lambda_1^3   &\alpha_1^*(\lambda_1^*)^3
                          &\alpha_2\lambda_2^3   &\alpha_2^*(\lambda_2^*)^3  
                           &\cdots&\alpha_q\lambda_q^3&\alpha_q^*(\lambda_q^*)^3
                            \\[2mm]
                      \vdots& \vdots&\vdots&               \vdots&               
                          \vdots   &\ddots&\vdots&\vdots\\[2mm]
                           1&\lambda_1^{2q}&(\lambda_1^*)^{2q}&\lambda_2^{2q}&
                           (\lambda_2^*)^{2q}&\cdots&\lambda_q^{2q}&
                           (\lambda_q^*)^{2q}\\[2mm] 
                    \end{array}\right|}
            {\left|\begin{array}{cccccccc} 
                           1&                1     &                         1&
                                           1     &                         1 
                                            &\cdots&1&                    1  
                                             \\[2mm]
                           1&\alpha_1\lambda_1     &\alpha_1^*\lambda_1^*   
                             &\alpha_2\lambda_2     &\alpha_2^*\lambda_2^*      
                              &\cdots&\alpha_q\lambda_q&\alpha_q^*\lambda_q^*  
                               \\[2mm]
                           1&\lambda_1^2           &(\lambda_1^*)^2&\lambda_2^2
                                      &(\lambda_2^*)^2             &\cdots&
                                      \lambda_q^2&(\lambda_q^*)^2         
                                       \\[2mm]
                           1&\alpha_1\lambda_1^3   &\alpha_1^*(\lambda_1^*)^3
                           &\alpha_2\lambda_2^3   &\alpha_2^*(\lambda_2^*)^3 
                             &\cdots&\alpha_q\lambda_q^3&\alpha_q^*
                             (\lambda_q^*)^3 \\[2mm]
                      \vdots& \vdots&\vdots&               \vdots&                 
                        \vdots   &\ddots&\vdots&\vdots\\[2mm]
                           1&\lambda_1^{2q}&(\lambda_1^*)^{2q}&\lambda_2^{2q}
                           &(\lambda_2^*)^{2q}&\cdots&\lambda_q^{2q}
                           &(\lambda_q^*)^{2q}\\[2mm] 
                    \end{array}\right|}
    \end{eqnarray}
    with (a bar denotes complex conjugation)
   \[\cdot\, f_0=\exp\left(-\int\limits_{-1}^1 \frac{(-1)^q g(x^2)\,dx}
   {Z_D}\right)
   ,\, Z_D=\sqrt{(\mbox{i}x-\zeta/\rho_0)^2+(\rho/\rho_0)^2}\,\,,\,\,
     \Re(Z_D)<0\]
    \[\cdot\,\lambda_\nu=\sqrt{\frac{Y_\nu-\mbox{i}\bar{z}}{Y_\nu+\mbox{i}z}},  
    \quad z=\frac{1}{\rho_0}(\rho+\mbox{i}\zeta),
    \quad \lambda_\nu^*\,\overline{\lambda}_\nu=1\]
    \[\cdot\,\alpha_\nu=\frac{1-\gamma_\nu}{1+\gamma_\nu},\quad 
    \gamma_\nu=\exp\left(\lambda_\nu(Y_\nu+\mbox{i}z)\int\limits_{-1}^1 
    \frac{(-1)^q g(x^2)\,dx}{(\mbox{i}x-Y_\nu)Z_D}\right),\quad 
    \alpha_\nu^*\,\overline{\alpha}_\nu=1\]
    satisfies the Ernst equation. With the additional requirement that for 
    each parameter $Y_\nu$
    there is also a parameter $Y_\mu$ with $Y_\nu=-\overline{Y}_\mu$, 
    reflectional 
    symmetry, $f(\rho,-\zeta)=\overline{f(\rho,\zeta)}$,
    is ensured\footnote{Hence, the set $\{\mbox{i}Y_\nu\}_q$ consists of 
    real parameters and/or pairs of complex conjugate parameters.}.
    Moreover, the parameters $Y_\nu$ are assumed to lie outside the imaginary 
    interval $[-\mbox{i},\mbox{i}]$.\\
    The above Ernst potential $f=f(\rho/\rho_0,\zeta/\rho_0;\{Y_\nu\}_q\,;g)$
    is obtained by a B\"{a}cklund transformation applied to the 
    real seed solution $f_0$, see \cite{N80a}. 
    On the other hand, as demonstrated in appendix A, it can be constructed 
    from the hyperelliptic solutions by a suitable limiting process 
    (see also \cite{Kor93}). 
    The particular ansatz chosen for the seed solution $f_0$ guarantees a 
    resulting Ernst potential which 
    corresponds to a disk-like source of the gravitational field (see also 
    section 1.2 of \cite{AM00}). \\
    Furthermore, $f$ does not possess singularities at 
    $(\rho,\zeta)=\rho_0(|\Im[Y_\nu]|,-\Re[Y_\nu])$. This is due to the fact 
    that $\alpha_\nu\lambda_\nu$ is a function 
    of $\lambda_\nu^2$, and this means that $f$ does not behave like a 
    square root function near
    the critical points  $(\rho,\zeta)=\rho_0(|\Im[Y_\nu]|,-\Re[Y_\nu])$, but 
    rather like a rational function. Now, in the 
    whole area of physically interesting solutions that will be treated in 
    the subsequent sections, 
    each zero of the denominator is cancelled by a corresponding zero of the 
    numerator in (\ref{G1.7}) 
    such that the resulting gravitational field is regular 
    outside the disk.\\[5mm]
    The real function $g$ that enters the Ernst potential is assumed to be 
    analytic on $[0,1]$ in order to guarantee an 
    analytic behaviour of the angular velocity $\Omega$ for all 
    $\rho\in[0,\rho_0]$. 
    Moreover, the additional requirement
    \[g(1)=0\] leads to a surface mass density $\sigma_p$ of the form
    \begin{equation}\label{sigma}
    \sigma_p(\rho)=\sigma_0\psi_p[(\rho/\rho_0)^2]\sqrt{1-(\rho/\rho_0)^2}
    \quad\mbox{[with $\psi_p$ analytic 
    in $[0,1],\,\psi_p(0)=1$]   }\end{equation}
    and therefore ensures that $\sigma_p$ vanishes at the 
    rim of the disk.  \\[5mm]
    In this article the question as to whether the above expression for the 
    Ernst potential is sufficiently
    general to approximate arbitrary differentially rotating disks of dust 
    is investigated. Of particular interest is a 
    rapidly converging method to perform this approximation. To this end, 
    the set $\{Y_\nu\}_q$ of complex parameters 
    will be translated into an analytic function \[\xi:[0,1]\to\mathbb{R}.\]
    Thus the Ernst potential will depend on two real analytic functions 
    defined on $[0,1]$:
    \[f=f(\rho/\rho_0,\zeta/\rho_0;\xi;g),\] which eventually proves to be 
    sufficient to satisfy both the 
    {\em dust condition} (\ref{G1.4}) and the 
    boundary condition (\ref{G1.3}) [or alternatively (\ref{G1.6})]. 
    The {\em rapid} and {\em accurate} approximation can be 
    realized since both $g$ and $\xi$ are analytic on $[0,1]$ and thus 
    permit elegant expansions in terms of 
    Chebyshev polynomials.
\section{Generalization of the B\"{a}cklund type solutions by a 
limiting process} 
As demonstrated in \cite{AM00} for the B\"{a}cklund type solutions 
with $q=1$, the {\em dust condition} (\ref{G1.4}) can be 
satisfied by an appropriate choice of the function $g$ if the complex 
parameters $Y_\nu$ are prescribed. To 
fulfil a second boundary condition, (\ref{G1.3}) or (\ref{G1.6}), 
the set $\{Y_\nu\}_q$ of these parameters has to be translated 
into a real analytic function $\xi$.\\
To this end, consider the following equalities for the above solutions 
$f=f(\{Y_\nu\}_q\,;g)$\footnote{In the following the 
Ernst potentials $f$ given by 
(\ref{G1.7}) are considered as complex functions depending on the set 
$\{Y_\nu\}_q$ of complex parameters and on $g$.} 
which are proved in appendix B:
    \begin{eqnarray}
        f[\{Y_1,\ldots,Y_{q-2},Y_{q-1},Y_{q}\};g]&=& 
        f[\{Y_1,\ldots,Y_{q-2}\};g]\label{G2.1}\\ &&\quad\mbox{if $\,
        Y_{q-1}=-Y_q\in\mathbb{R}$}\nonumber\\[1mm]
        f[\{Y_1,\ldots,Y_{q-2},Y_{q-1},Y_{q}\};g]&=& 
        f[\{Y_1,\ldots,Y_{q-2}\};g]\label{G2.2}\\&&\quad\mbox{if }\, 
        Y_{q-1}=\overline{Y}_q\nonumber\\[1mm]
        \lim\limits_{t\to \infty }  f[\{Y_1,\ldots,
        Y_{q-1},\mbox{i}t\};g] 
                      &=&f[\{Y_1,\ldots,Y_{q-1}\};g]\label{G2.2a}\\[-1mm]&&
                     \quad \mbox{if $\,t\in\mathbb{R}$}\nonumber\\[1mm]
        \lim\limits_{Y_q\to \infty }  f[\{Y_1,\ldots,
        Y_{q-2},Y_{q-1},Y_q\};g] 
                      &=&f[\{Y_1,\ldots,Y_{q-2}\};g]\label{G2.2b}\\[-2mm]&&
                 \quad\mbox{if $\,Y_{q-1}=-\overline{Y}_q\;.$}\nonumber
        \end{eqnarray}
In order to find an approximation scheme, the desired function 
$\xi=\xi(\{Y_\nu\}_q)$ is supposed to be invariant 
under the modifications (\ref{G2.1}-\ref{G2.2b}) of the set 
$\{Y_\nu\}_q$ that do not effect the Ernst potential. This property will be 
necessary to 
solve the boundary conditions uniquely. \\
It is realized by the real analytic function
\begin{equation}\label{G2.3}
\xi(x^2;\{Y_\nu\}_q)=\frac{1}{x}\ln\left[\,\prod_{\nu=1}^q
\frac{\mbox{i}\,Y_\nu-x}{\mbox{i}\,Y_\nu+x}\right],\quad x
\in[-1,1],\end{equation}
which can be proved by considering that for each parameter $Y_\nu$ 
there is also a parameter $Y_\mu$ with 
$Y_\nu=-\overline{Y}_\mu\,$, and that, moreover, the parameters $Y_\nu$ 
do not lie on the imaginary interval 
$[-\mbox{i},\mbox{i}]$.\\[5mm]
The set ${\cal X}$ of all functions $\xi=\xi(x^2;\{Y_\nu\}_{q}),\,q\in
\mathbb{N}$, which are defined by (\ref{G2.3}) forms a dense subset of 
the set ${\cal A}$ of all real analytic functions on $[0,1]$. Now, for a 
given function $g$, each $\xi\in{\cal X}$ is  
mapped by (\ref{G1.7}) onto a uniquely defined Ernst potential $f\in
{\cal E}$ \footnote{Here, ${\cal E}$ denotes the set of 
all Ernst potentials corresponding to disk-like sources.} :
\begin{equation}\label{Phi}\Phi_g:{\cal X}\longrightarrow{\cal E},\quad 
\Phi_g(\xi)=f(\{Y_\nu\}_q;g),\end{equation}
where the set $\{Y_\nu\}_q$ results from $\xi$ by (\ref{G2.3}).\\
In the following, it is assumed that this mapping $\Phi_g$ can be 
extended to form a continuous function defined on ${\cal
A}$.\footnote{The mathematical aspects of this assumption will be 
discussed in section 5.}
Then, given the two real functions $g$ and $\xi$, defined and 
analytic on the interval $[0,1]$, the Ernst potential 
\[f(\xi;g)=\lim_{q\to\infty}f(\{Y_\nu^{(q)}\}_q;g)\] exists and is 
independent of the particular choice of the sequence
$\{\{Y_\nu^{(q)}\}_q\}_{q=q_0}^\infty$ which serves to represent $\xi$ by
\begin{eqnarray}\label{G2.4} 
\xi(x^2)=\frac{1}{x}\lim_{q\to\infty}\,
     \ln\left[\,\prod_{\nu=1}^q\frac{\mbox{i}\,Y_\nu^{(q)}-x}{\mbox{i}
     \,Y_\nu^{(q)}+x}\right]
\quad \mbox{ for}\quad x\in[-1,1].    \nonumber
\end{eqnarray}
This provides the groundwork for the approximation scheme that will be 
developed in the next section.
The treatment additionally assumes that the boundary conditions (\ref{G1.3}) 
and 
(\ref{G1.4}) [or (\ref{G1.4}) and (\ref{G1.6})] interpreted as functions of 
$g$ and $\xi$ are invertible. The 
accurate and rapid convergence of the numerical methods justifies this 
assumption although a rigorous proof cannot be 
given.
\section{An approximation scheme for arbitrary differentially rotating 
disks of dust}
It is now possible to attack general boundary value 
problems for differentially rotating disks of dust.
With the above generalized solutions $f=f(\xi;g)$ the boundary conditions 
[see formulas (\ref{G1.3}-\ref{G1.6}, \ref{sigma})]
become a problem of inversion to determine $g$ and $\xi$ 
from $\sigma_p$ or $\Omega$:
 \\[5mm]\parbox{10cm}{
\begin{tabular}{cl} 
   {\em (A)} & $S(g;\xi)=\{e^{U-k}\,[U_{,\zeta}+\frac{1}{2}Q]/
   [\sigma_0\sqrt{1-(\rho/\rho_0)^2}]\}(\xi;g)
               \doteq2\pi\psi_p\quad\mbox{or}$\\[5mm] 
   {\em (A')}& $O(g;\xi)=\{Q/[\Omega(0)(a_{,\zeta}-a\;Q)]\}(\xi;g)
   \doteq\Omega/\Omega(0)=\Omega^*$\\[5mm]
   {\em (B)} & $D(g;\xi)=\left\{\rho_0^2\left[Q^2e^{4U}+Q\,
   (e^{4U})_{,\zeta}+(b_{,\rho})^2\right]\right\}(\xi;g)\doteq0,
   \quad g(1)\doteq0$
\end{tabular}
}\hfill
\parbox{1cm}{\begin{eqnarray} \label{AS1}\end{eqnarray}}\\[5mm]
This inversion problem is tackled in the following manner:
\begin{enumerate}
\item The only way to treat the complicated system (\ref{AS1}) numerically 
seems to be by restricting it to a finite, 
discretized version and solving this by means of a Newton-Raphson method. 
\item For this method, a good initial guess for the 
solution is needed. As shown in appendix C.1, there exists 
a representation of the functions $g$ and $\xi$ in 
terms of $\sigma_p$ or $\Omega$ in the Newtonian regime $\varepsilon\ll 1$ 
where $\varepsilon=M^2/J$ and the 
gravitational mass $M$ and the total angular momentum $J$ are given by
\begin{equation}\label{AS2}
M=2\int\limits_S(T_{ab}-\frac{1}{2}Tg_{ab})n^a\xi^b dV\,,\;
J=-\int\limits_ST_{ab}n^a\eta^bdV\,,\;T_{ab}=g_{ab}T^{ab}.
\end{equation}
($S$ is the spacelike hypersurface $t=constant$ with the unit  future-pointing
 normal vector $n^a\,$; the Killingvectors 
$\xi^a$ and $\eta^a$ correspond to stationarity and axisymmetry, 
respectively.)
\item This motivates the following finite version which results from 
expansions of (\ref{AS1}) in terms of 
Chebyshev-polynomials $T_j(\tau)=\cos[j\arccos(\tau)]$:
\begin{eqnarray}&&\hspace*{-5mm}F_j(v_k)\doteq0\quad 
(1\leq j,k\leq N_1+N_2-1):\nonumber\\ [5mm]
   &&\nonumber\begin{array}{l}
   \cdot\, F_j=D_j\quad(1\leq j\leq N_1-1),\quad 
   F_{N_1}=\varepsilon(g_m;\xi_n)-\varepsilon,\\[5mm]
   \quad  F_{N_1+j-1}=S_j-2\pi\psi_j\quad\mbox{or}\quad 
   F_{N_1+j-1}=O_j-\Omega^*_j \quad(2\leq j\leq N_2),\\[5mm]
   \quad v_k=g_{k+1}\quad(1\leq k\leq N_1-1),\quad 
   v_{N_1+k-1}=\xi_k\quad(1\leq k\leq N_2)\\[5mm]
   \cdot\,g(x^2)\approx\sum\limits_{j=1}^{N_1} 
   g_j T_{j-1}(2x^2-1)
          -\frac{1}{2}g_1,\quad g(1)\doteq 0\,\Rightarrow\,
          g_1=-2\sum\limits_{j=2}^{N_1} g_j\\[5mm]
   \cdot\,\xi(x^2)\approx\sum\limits_{j=1}^{N_2} \xi_j T_{j-1}(2x^2-1)
   -\frac{1}{2}\xi_1\\[5mm]
   \cdot\,\psi_p(x^2)\approx\sum\limits_{j=1}^{N_2} \psi_j T_{j-1}(2x^2-1)
   -\frac{1}{2}\psi_1,\\[5mm]
   \quad\psi_p(0)\doteq 1\,\Rightarrow\,\psi_1=2\sum\limits_{j=2}^{N_2} 
   (-1)^j\psi_j+2\\[5mm]
   \cdot\,\Omega^*[(\rho/\rho_0)^2]=\Omega(\rho)/\Omega(0):\\[3mm]
   \quad\Omega^*(x^2)\approx\sum\limits_{j=1}^{N_2} \Omega^*_j 
   T_{j-1}(2x^2-1)-\frac{1}{2}\Omega^*_1,\\[3mm]
         \quad \Omega^*(0)\doteq 1\,\Rightarrow\,\Omega^*_1=
         2\sum\limits_{j=2}^{N_2} (-1)^j\Omega^*_j+2\\[5mm]
   \cdot\,S(x^2=\rho^2/\rho_0^2\,;g;\xi)\approx\sum\limits_{j=1}^{N_2} 
   S_j(g_m;\xi_n) 
          T_{j-1}(2x^2-1)-\frac{1}{2}S_1(g_m;\xi_n)\\[5mm]
   \cdot\,O(x^2=\rho^2/\rho_0^2\,;g;\xi)\approx\sum\limits_{j=1}^{N_2} 
   O_j(g_m;\xi_n) 
          T_{j-1}(2x^2-1)-\frac{1}{2}O_1(g_m;\xi_n)\\[5mm]
   \cdot\,D(x^2=\rho^2/\rho_0^2\,;g;\xi)\approx\sum\limits_{j=1}^{N_1-1} 
   D_j(g_m;\xi_n) T_{j-1}(2x^2-1)-\frac{1}{2}D_1(g_m;\xi_n)
\end{array}\end{eqnarray}
[The function $\varepsilon(g_m;\xi_n)=M^2/J$ is determined using 
(\ref{AS2}) for the above functions $g$ and $\xi$.]
\item For the above system, the boundary values are assumed to be given in 
the form of the $\psi_k$'s or 
$\Omega^*_k$'s $\,(k=2,\ldots,N_2)$. Moreover, some $\varepsilon\ll 1$ has 
to be prescribed. Then, good initial $v_k$'s come 
from the Newtonian 
expansion. The Newton-Raphson method improves the $v_k$'s and yields a very 
accurate solution to (\ref{AS1}) for the 
chosen small $\varepsilon$. Now, this solution serves as the initial estimate 
for the $v_k$'s belonging to a marginally increased 
value for $\varepsilon$. Again, the Newton-Raphson method improves the 
solution, and one continues in this manner until 
this procedure ceases to converge. This occurs for some finite value 
$\varepsilon_0$, at the latest for $\varepsilon=1$. 
A further discussion of this limit is given below.
\item A rather technical detail is the retranslation of the $\xi_j$ 
into a set $\{Y_\nu\}_q$ 
which then gives a satisfactory approximation of $\xi$ in terms of 
(\ref{G2.3}). There are many ways 
to do this. Here, the following one has been chosen.\\
One rewrites equation (\ref{G2.3}) in the equivalent form
\[\exp\left[x\,\xi(x^2;\{Y_\nu\}_q)\right]=
    \prod_{\nu=1}^q\frac{\mbox{i}\,Y_\nu-x}{\mbox{i}\,Y_\nu+x}=
    \frac{P_{q}(-x)}{P_{q}(x)}\quad\mbox{with}\quad P_{q}(x)=
    \sum_{\nu=1}^{q}b_\nu x^\nu.\]
The coefficients $b_\nu$ of the polynomial $P_{q}$ can be determined 
by evaluating the left hand 
side at $q$ arbitrary different points $x_\mu\in[0,1]$ \footnote{Here, 
zeros of Chebyshev-polynomials have been used.} and solving the following 
linear system: 
\[\exp\left[x_\mu\,\xi(x_\mu^2;\{Y_\nu\}_q)\right]\sum_{\nu=1}^{q}
b_\nu x_\mu^\nu=
     \sum_{\nu=1}^{q}b_\nu (-x_\mu)^\nu\]
The zeros of $P_{q}$ determine the $Y_\nu$.
\end{enumerate}
The above scheme has been performed for many different prescribed surface 
mass densities and angular 
velocities. This provides strong evidence for the conjecture that, in this 
manner, all Newtonian disks 
can be extended into the relativistic regime.
It has been found that the value for $\varepsilon_0$, the limiting parameter
for the convergence of this scheme, depends on the chosen profile for 
$\psi_p$ (or equivalently for $\Omega^*$). 
It is illustrated in appendix C.2, how the Ernst potential always tends to 
the extreme 
Kerr solution \cite{Kerr63} as $\varepsilon\to 1$. This supports a 
conjecture by Bardeen and Wagoner \cite{BW}.
But $\varepsilon_0=1$ does not hold for all given surface mass densities. 
Even in the Newtonian regime there are 
surface mass densities for which a realistic physical disk cannot be found 
since the corresponding angular 
velocity would become imaginary. If one chooses a profile for $\sigma_p$ not 
very different from these, then the 
Newtonian limit still might exist, but some $\varepsilon_0<1$ turns up, 
beyond which 
the method does not converge. In the case of prescribed angular velocity, 
the situation is similar. Here, for any sequence 
$f=f(g_\varepsilon;\xi_\varepsilon)$ the angular velocity $\Omega^*$ tends 
for all 
$x^2\in[0,1]$ to 1 as $\varepsilon\to 1.$
So, each nonuniform rotation law will lead to some
$\varepsilon_0<1$ (see section 4 for examples).\\[5mm]
The above expansions in terms of Chebyshev-polynomials allow a very accurate 
representation with only a small number of 
coefficients. However, the retranslation of $\xi$ (see the above point 5) 
leads to functions that are not especially 
well suited for an approximation. In particular, if the boundary condition 
$\psi_p$ is chosen to be close to those 
for which there is no Newtonian disk, then the accuracy cannot be driven 
particularly high by the computer 
program used, although the method in priniciple allows arbitrary 
approximation (see section 4.2). \\
For $\psi_p$'s sufficiently far away from those critical ones, 
the accuracy obtained 
was very high. By choosing appropriate values for $N_1$ and $N_2$ one 
can always achieve  extremly good agreement with
the {\em dust condition} (\ref{G1.4}) (12 digits and beyond) which ensures 
a realistic physical interpretation of the solution.
The accuracy to which the second boundary condition, (\ref{G1.3}) 
or (\ref{G1.6}), can be satisfied, depends on the 
parameter $\varepsilon$. It is usually around 8 digits in the weak 
relativistic regime, and falls as 
$\varepsilon$ increases, but is still around 4 digits as $\varepsilon$ 
tends to $\varepsilon_0$. These values arose for 
$N_1=30,\,N_2=12,$ and typical $\psi_p$'s (like $\psi_p$'s depending 
linearly on $x^2$) and $\Omega^*$'s (e.~g.~the 
realistic one considered in section 4.1). The number $q$ of the 
parameters $Y_\nu$ by which $\xi$ is represented, was chosen to be 
between 20 and 30 (independently of $N_2$). \\[5mm]
What remains to be discussed is the regularity of the 
Ernst potentials that were obtained. For a few of the solutions, the 
functions $e^{2U}$ and $b$ were plotted over the coordinates 
$\rho$ and $\zeta$. Moreover, 
the agreement of the alternative representations of $M$ and $J$, 
as given by the behaviour of the 
Ernst potential at infinity
\[U=-\frac{M}{r} +{\cal O}(r^{-2}),\quad b=-2J\,\frac{\cos\theta}{r^2}+
{\cal 
O}(r^{-3}),\quad(r=\sqrt{\rho^2+\zeta^2},\,\zeta=r\cos\theta)\]
with the results from formulas (\ref{AS2}) yields good confirmation of 
the regularity. This agreement was checked for all solutions that were 
calculated.
\section{Representative examples}
From the numerous solutions obtained, three particular sets of 
differentially rotating disks are discussed in more detail. The first 
one is an example of disks revolving with a 
realistic rotation law. The second set illustrates the break down 
of the numerical method for a specially prescribed 
surface mass density $\sigma_p$ at some $\varepsilon_0<1$. On the other 
hand it is demonstrated that, for the same 
$\sigma_p$, regular solutions can be found in the highly relativistic regime. 
Finally, the third example concerns the 
occurence of a second ergoregion for a particular series of disks and, 
moreover, the gradual merging of the two spatially 
separated ergoregions as $\varepsilon$ increases.\\
The deviations between the boundary values obtained for particular numerical 
solutions and the given boundary conditions 
are listed in tables. The quantities $\Delta_D,\Delta_\Omega$, and 
$\Delta_\sigma$ therein are defined by
\[\Delta_D=\max_{x^2\in[0,1]}|D_{\mbox{\footnotesize obt}}(x^2;g;\xi)|\]
\[\Delta_\Omega=\max_{x^2\in[0,1]}\left|\Omega_{\mbox{\footnotesize obt}}^*
(x^2)
-\Omega_{\mbox{\footnotesize giv}}^*(x^2)\right|\]
\[\Delta_\sigma=\max_{x^2\in[0,1]}\left|\psi_p^{\mbox{\footnotesize obt}}
(x^2)
-\psi_p^{\mbox{\footnotesize giv}}(x^2)\right|,\]
where the indices 'obt' and 'giv' refer to obtained and given quantities, 
respectively. Moreover, by letters 
(a)$,\ldots,$(e), special examples are marked, for which illustrative graphs 
have been made. Here, curves drawn 
in the same line style belong to the same solution. The graphs show the 
dimensionless quantities 
$\rho_0\sigma_p$ and $\rho_0\Omega$ as well as $g$ and $\xi$ plotted 
against the normalized 
radial coordinate $\rho/\rho_0$ and $x$, respectively.
\subsection{Disks possessing a realistic rotation law}
As motivated by observations in astrophysics the rotation law of a 
galaxy is often modelled by an equation of the form
(see \cite{Brandt})
\begin{equation}\label{Om} \Omega(\rho)=
\frac{\Omega(0)}{\sqrt{1+\rho^2/\rho_1^2}}.\end{equation}
Here, the parameter $\rho_1$ varies for different galaxies. 
In the following series of solutions 
illustrated in figure 1, $\rho_1=0.7\rho_0$ has been chosen. 
As described in section 3, there is a limiting parameter 
$\varepsilon_0\approx 0.935$, 
for which the numerical method ceases to converge.\\
\begin{center}
\begin{figure}[h]
\unitlength1cm
\begin{minipage}[h]{20cm}
\begin{picture}(500,12)
\put(0.18,-0.5){\epsfig{file=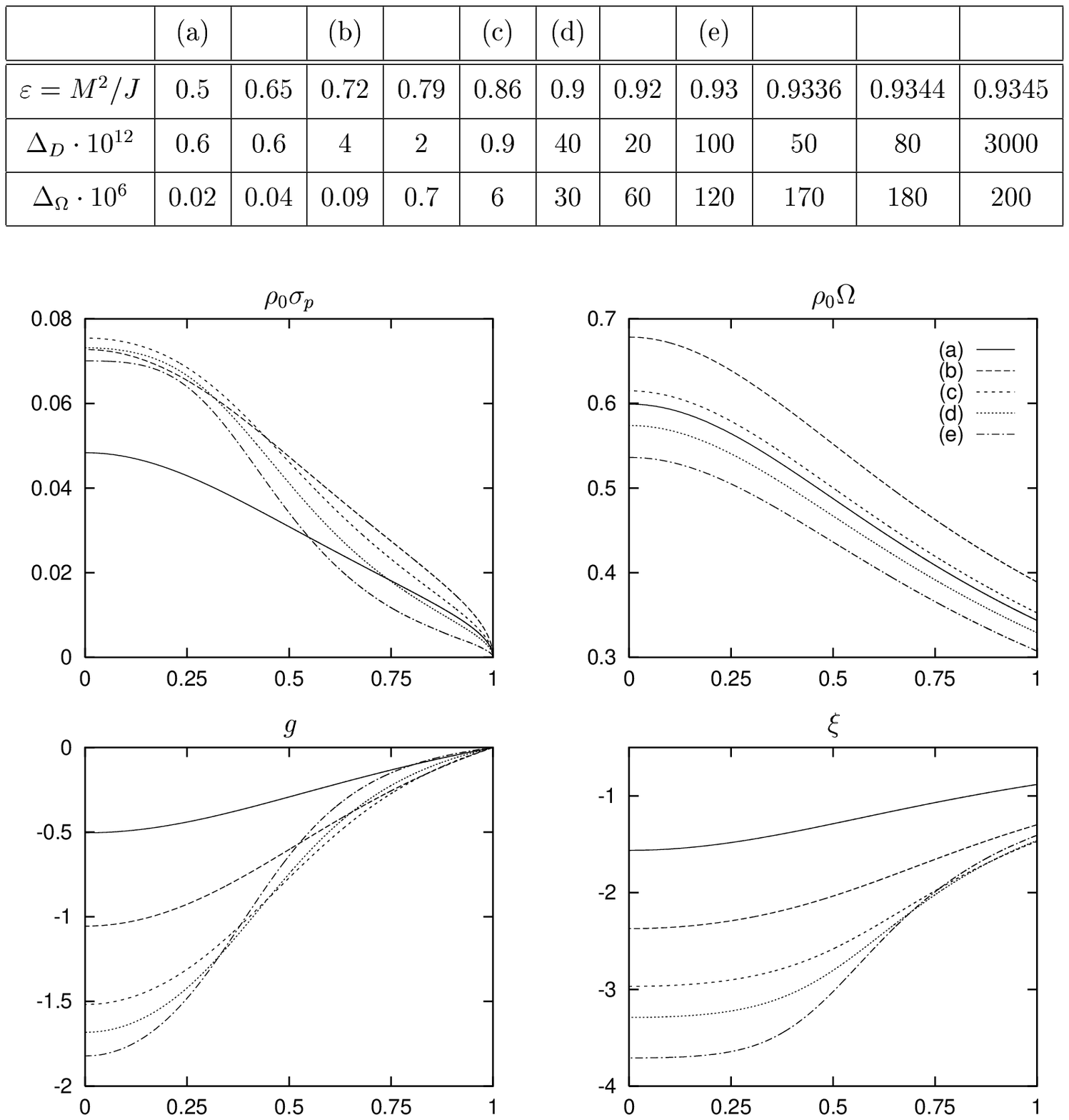,scale=0.78}}
\end{picture}
\end{minipage}
\end{figure}
\vspace*{0.8cm}
Figure 1: Disks possessing the rotation law (\ref{Om}) with 
$\rho_1=0.7\rho_0$ $(N_1=30,N_2=12)$.
\end{center}
\subsection{Disks with a critical surface mass density}
For the following sequence of solutions, a surface mass density of the form
\begin{equation}\label{sig} \sigma_p(\rho)=\sigma_0\left(1-3\,\frac{\rho^2}
{\rho_0^2}+\beta\,\frac{\rho^4}{\rho_0^4}\right)
                            \sqrt{1-\frac{\rho^2}{\rho_0^2}}\end{equation}
has been assumed.\\
\begin{center}
\begin{figure}[h]
\unitlength1cm
\begin{minipage}[h]{20cm}
\begin{picture}(500,12)
\put(0.18,-0.5){\epsfig{file=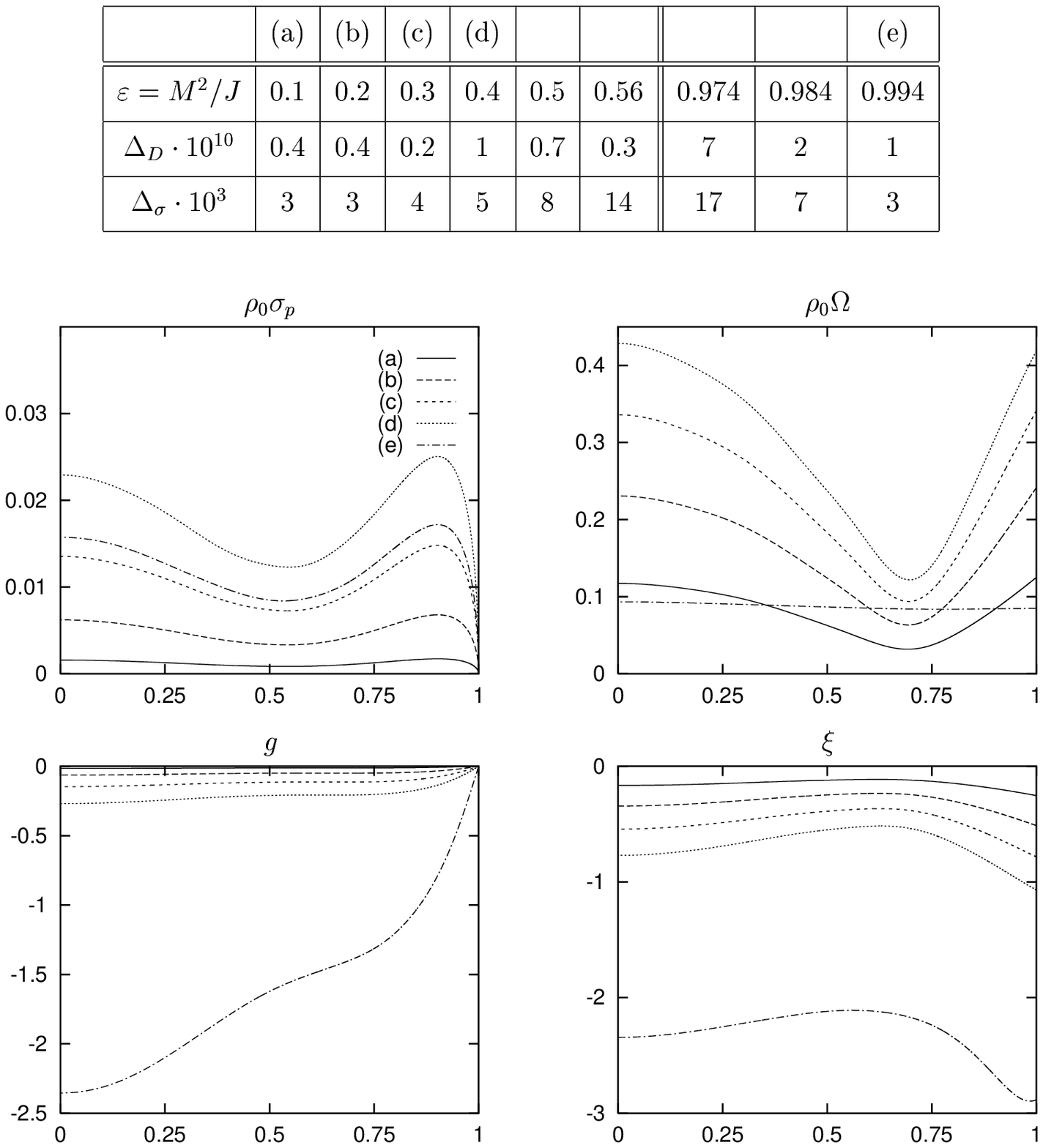,scale=0.78}}
\end{picture}
\end{minipage}
\end{figure}
\vspace*{0.3cm}
Figure 2: Disks possessing the surface mass density (\ref{sig}) 
with $\beta=6$ $(N_1=30,N_2=12)$.
\end{center}
It turns out that for $\beta>\beta_N\approx 7$ no Newtonian disks with 
a real angular velocity can be 
found. On the other hand, for $\beta=5.5$, all relativistic solutions 
for $0\leq\varepsilon\leq 1$ exist. The table and 
graphs of figure 2 refer to the case $\beta=6$. 
Starting here from the Newtonian solution, one soon recognizes a first 
limiting parameter
$\varepsilon_0\approx 0.60$ for which the method breaks down. However, 
by coming from solutions with $\beta=5.5$ and 
$\varepsilon$ close to 1, it is possible to create highly relativistic 
solutions with $\beta=6$. In fact, there is 
another limiting parameter, $\varepsilon_1\approx0.97$, above which the 
solutions with $\beta=6$ exist once again. Due to 
the nearness to the critical surface mass density (for $\beta=\beta_N$), 
the accuracy obtained for the boundary condition 
(\ref{G1.3}) is not very high.
\subsection{Disks possessing spatially separated ergoregions}
The particular set of disks depicted in figure 3 demonstrates 
the occurence of a second 
ergoregion\footnote{An ergoregion is a portion of the $(\rho,\zeta)$-space 
within which the function $e^{2U}$ is negative.}.
These solutions do not satisfy a specially prescribed boundary condition 
(\ref{G1.3}) or (\ref{G1.6}), but have been
constructed in the following manner as intermediate solutions. 
\newpage
\begin{center}
\begin{tabular}       {|c|c|c|c|c|c|}\hline
{\rule[-3mm]{0mm}{8mm}}&(a)&(b)&(c)&(d)&(e)\\ \hline\hline
{\rule[-3mm]{0mm}{8mm}}$\varepsilon=M^2/J$&0.84038&0.84054&
0.84079&0.84120&0.84162\\ \hline
{\rule[-3mm]{0mm}{8mm}}$\Delta_D\cdot 10^{12}$&3&2&2&5&2\\ \hline
\end{tabular}
\end{center}
\vspace*{5mm}

\begin{center}
\begin{figure}[h]
\unitlength1cm
\begin{minipage}[h]{20cm}
\begin{picture}(500,12)
\put(0.25,-3.5){\epsfig{file=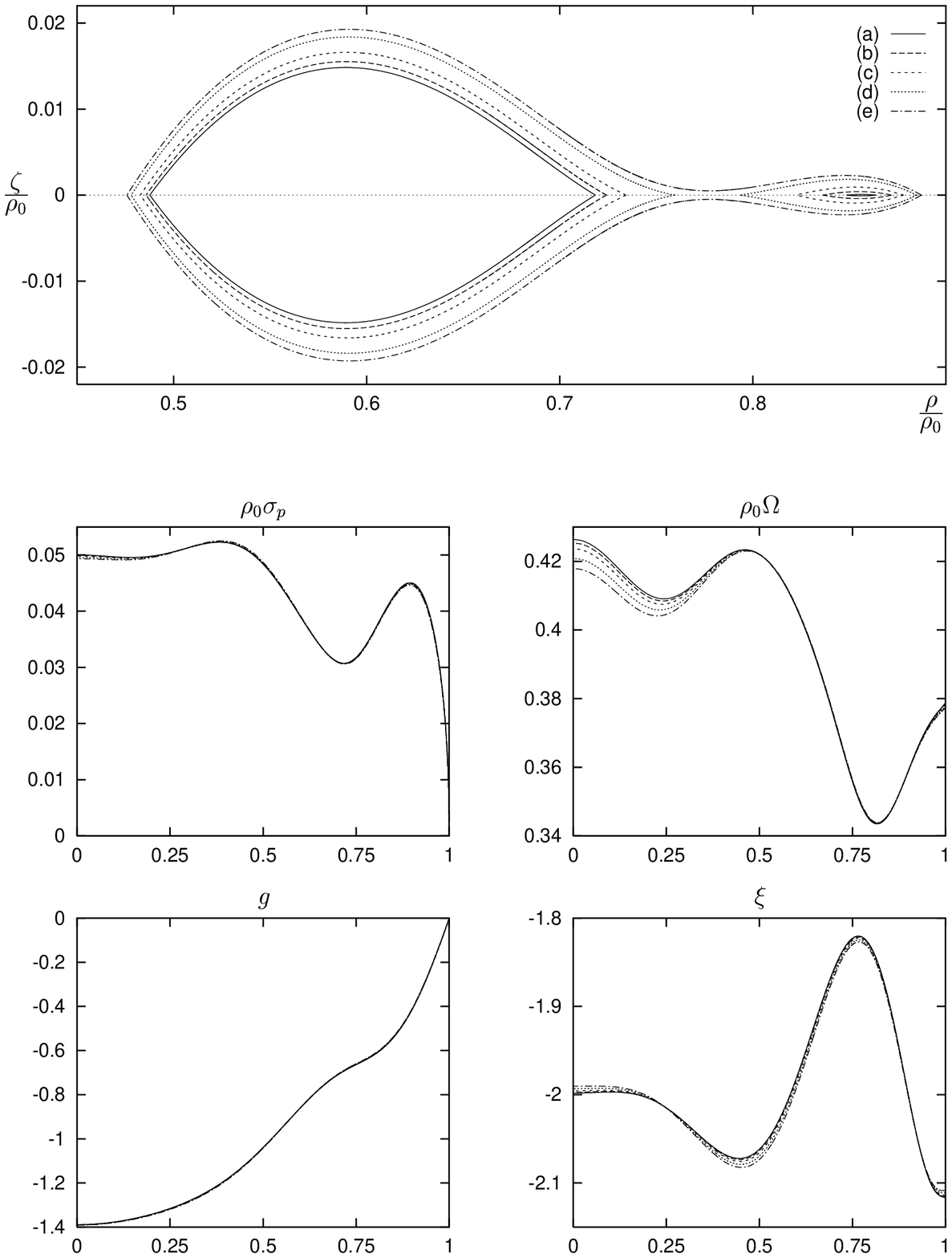,scale=0.75}}
\end{picture}
\end{minipage}
\end{figure}
\end{center}
\vspace*{3.5cm}
Figure 3: Example for a series of disks possessing spatially separated 
ergoregions. In the uppermost picture, 
the rims of the ergoregions in the $(\rho/\rho_0,\zeta/\rho_0)$-space are 
to be seen $(N_1=40,N_2=9)$. 
\newpage
If one investigates solutions with surface 
mass densities similar to those of (\ref{sig}), one recognizes two minima 
for $e^{2U}$ (taken as a function of 
$\rho,\,0\leq\rho\leq\rho_0,\,\zeta=0$), say at $\rho_a$ and $\rho_b>\rho_a$. 
Now, for a particular choice of $\sigma_p$ it is possible to get 
$e^{2U}(\rho_a)>0$ and $e^{2U}(\rho_b)<0$, whilst 
by another choice one can achieve $e^{2U}(\rho_a)<0$ and $e^{2U}(\rho_b)>0$. 
This makes clear, that disks with spatially 
separated ergoregions can be constructed by interpolating between these 
solutions.\\
For the chosen example, there is only a narrow interval $(\varepsilon_a,
\varepsilon_b)$ for which the two separated ergoregions occur. 
As can be seen from figure 3, after creation of the second ergoregion at 
$\varepsilon_a\approx0.8403$, 
both ergoregions grow as $\varepsilon$ increases. Eventually, at 
$\varepsilon_b\approx0.8415$, the ergoregions merge into 
one ergoregion. 

\section{Discussion of mathematical aspects}
As already mentioned in section 2, the assumption that the function $\Phi_g$
introduced in (\ref{Phi}) can be extended to form a continuous
mapping defined on $\cal A$, lies at the heart of the above numerical methods.
Although this assumption 
seems to be intuitive, it is not trivial. Consider the following example:\\
For any analytic function $\psi:[0,1]\to\mathbb{R}$ one finds the
equality\footnote{To verify this formula one 
simply expands the logarithms in the form $\ln(1+\epsilon)=\epsilon+{\cal
O}(\epsilon^2)$ and notes that the resulting 
sum tends to the Riemann integral of the right hand side.}:
\[\lim_{q\to\infty}\sum_{\nu=1}^q\ln\left[1+\frac{1}{q}\psi\left(\frac{\nu}{q}
\right)\right]=\int\limits_0^1\psi(t)\,dt.\]
From this it follows that 
\[2\int_0^1\phi(t)\,dt=\frac{1}{x}\lim_{q\to\infty}\,\sum_{\nu=1}^q\ln
\frac{q+x\phi(\nu/q)}
                                                           {q-x\phi(\nu/q)}
                                                           \qquad\mbox{with
                                                           $\psi(t)=\pm\,
                                                           x\,\phi(t)$.}\]

Hence, the function $\xi(x^2)\equiv 2$ can be represented by any sequence of 
the form
\[Y^{(q)}_\nu=
\mbox{i}\frac{q}{\phi(\nu/q)}\quad\mbox{with}\quad\int_0^1\phi(t)\,dt=1.\]
Since these sequences might be quite different from each other, it is 
rather surprising that all of them approximate the same Ernst potential given 
by (\ref{G1.7}). But this follows from the above
assumption.\\
This already indicates the difficulties which are connected with a rigorous
proof of this assumption because the Ernst 
potential is only given in terms of the set $\{Y_\nu\}_q$ and not directly in
terms of $\xi$.\\[5mm]
A further conjecture is strongly confirmed by extensive numerical 
investigations:
\begin{quote}{\em  
For the hyperelliptic class of solutions represented by (\ref{AG0}) in appendix
A, the functions $\xi$ and $g$ are given by 
\begin{eqnarray}
    \xi(x^2)&=&\frac{1}{2x}\ln\left[\,\prod\limits_{\nu=1}^p\frac{\mbox{\em
    i}X_\nu-x}{\mbox{\em i}X_\nu+x}\right]\nonumber\\
    g(x^2)&=&\mbox{\em sign}\left(\,\prod_{\nu=1}^p X_\nu\right) A_g(x^2)
    h(x^2)\,,\nonumber\\
    &&\qquad\quad A_g(x^2)=\sqrt{\prod\limits_{\nu=1}^p(\mbox{\em
    i}x-X_\nu)(\mbox{\em i}x-\overline{X}_\nu)}\,,\quad A_g(x^2)>0. 
    \nonumber
\end{eqnarray}
In particular, in this formulation, the solution for the Neugebauer-Meinel-disk
\cite{NM93,NM95,NKM96} assumes the form $f=f(\xi;g)$
where
\begin{eqnarray}
    \xi(x^2)&=&\frac{1}{2x}\ln\frac{x^2-C_1(\mu)x+C_2(\mu)}
    {x^2+C_1(\mu)x+C_2(\mu)}\,,\nonumber\\
        &&\qquad\quad C_1(\mu)=\sqrt{2[1+C_2(\mu)]}\,,\quad
    C_2(\mu)=\frac{1}{\mu}\sqrt{1+\mu^2}\,,\nonumber\\
    g(x^2)&=&-\frac{1}{\pi}\,\mbox{\em arsinh}[\mu(1-x^2)]\,,  \nonumber
\end{eqnarray}
and the parameter $\mu,\,0<\mu<\mu_0=4.62966184..,$ is related to the angular
velocity by
\[   \mu=2\Omega^2\rho_0^2e^{-2V_0} ,\quad V_0=U(\rho=0,\zeta=0).\]
}
\end{quote}
As already mentioned, a direct proof of the above assumptions promises to 
be very complicated. But there might be an 
alternative proof which relies on relating a general solution of the Ernst 
equation to the solution  of a so-called 
Riemann-Hilbert problem, see \cite{Hausera,N96,N00,Aleks99}. In this treatment,
an appropriately introduced matrix function, from which 
the Ernst potential can be extracted, is supposed to be regular on a 
two-sheeted Riemann surface of genus zero except 
for some given curve, where it possesses a well-defined jump behaviour.
The freedom of two jump functions defined on this curve corresponds to the 
freedom to choose $\xi$ and $g$. 
Now, if one succeeds in finding a particular formulation of a Riemann-Hilbert 
problem in which $\xi$ and $g$ are involved, 
then the final solution for $f$ proves to depend only on $\xi$ (and $g$)
 and not on a particular global representation in terms of $\{Y_\nu\}_q$. 
 This deserves 
further investigation.\\[5mm]
There is very strong numerical evidence for the validity of both assumptions. 
For various functions $\xi$ (and functions 
$g$), different representations $\{Y_\nu\}_q$ have been seen to approximate 
the same Ernst potential. In particular, 
the approximation of the Neugebauer-Meinel-solution in terms of B\"{a}cklund 
solutions was carried out to give an 
agreement up to the 12th digit with the hyperelliptic solution, which 
confirms both assumptions. 
\begin{appendix}
\section{The transition from the hyperelliptic solutions to the B\"{a}cklund 
type solutions}
In this section the B\"{a}cklund type solutions are derived from the 
hyperelliptic class. 
The latter is assumed to be 
given in the form represented in \cite{AM00}\footnote{The parameters 
$K_\nu$, the upper 
integration limits $K^{(\nu)}$, and the integration variable $K$ have to be 
replaced by 
their 'normalized' values $X_\nu=K_\nu/\rho_0,\,X^{(\nu)}=K^{(\nu)}/\rho_0$, 
and $X=K/\rho_0$, respectively.} for an even integer $p\geq 2$:
\begin{equation}\label{AG0}f=\exp\left(\sum_{\nu=1}^p
\int\limits_{X_\nu}^{X^{(\nu)}}\frac{X^pdX}{V(X)}-u_p\right)
\end{equation}
\[\quad\cdot\,V(X)=\sqrt{(X+\mbox{i}z)(X-\mbox{i}\bar{z})
\prod_{\nu=1}^p(X-X_\nu)(X-\bar{X}_\nu)}\,\,,\quad 
z=\frac{1}{\rho_0}(\rho+\mbox{i}\zeta)\]
\begin{equation}\label{AG1}\quad\cdot\,\sum_{\nu=1}^p\int
\limits_{X_\nu}^{X^{(\nu)}}\frac{X^jdX}{V(X)}=u_j,   \quad 0\leq j<p 
\end{equation}
\[\quad\cdot\,u_j=\int\limits_{-1}^1 \frac{(\mbox{i}x)^j 
h(x^2)\,dx}{Z_D},\quad 0\leq j\leq p,\quad h:[0,1]\to 
\mathbb{R},\,\mbox{analytic},\]
\[\hspace*{8cm}\mbox{$Z_D$ as defined in (\ref{G1.7})} \]
The set $\{\mbox{i}X_\nu\}_p$ consists of arbitrary real parameters 
and/or pairs of complex conjugate parameters (in order 
to guarantee reflectional symmetry). The ($z$-dependent) values for the 
$X^{(\nu)}$ as well as the 
integration paths on a two-sheeted Riemann surface result from the Jacobian 
inversion problem (\ref{AG1}).\\
The transition to the B\"{a}cklund type solutions (\ref{G1.7}) can be 
obtained in the limit $\varepsilon\to 0$ by the following 
assumptions:
\begin{itemize}
\item $p=2q$
\item $X_{2\nu-1}=Y_\nu+\varepsilon\beta_\nu,\quad X_{2\nu}=
Y_\nu\quad(1\leq\nu\leq q),\quad\{\beta_\nu\}_q\mbox{ arbitrary}$
\item $g(x^2)=(-1)^q h(x^2)A(\mbox{i}x),\quad
      A(X)=\prod\limits_{\nu=1}^q(X-Y_\nu)(X-\overline{Y}_\nu)$ .\\
\end{itemize}
To this end, the above expression for $f$ is rewritten in the equivalent form:
\[f=\exp\left[\sum_{\nu=1}^q\left(\int_{X_{2\nu-1}}^{X^{(2\nu-1)}}\frac{A(X)dX}
{V(X)}
                                 +\int_{X_{2\nu}}^{X^{(2\nu)}}\frac{A(X)dX}
                                 {V(X)}\right)-
                                 \int\limits_{-1}^1 \frac{(-1)^q g(x^2)
                                 \,dx}{Z_D}\right]\]
The Jacobian inversion problem (\ref{AG1}) reads as follows in a similarly 
rewritten form $(1\leq\mu\leq q)$:
\[\cdot\,\sum_{\nu=1}^q\left(\int_{X_{2\nu-1}}^{X^{(2\nu-1)}}\frac{A(X)dX}
{V(X)(X-Y_\mu)}
                                 +\int_{X_{2\nu}}^{X^{(2\nu)}}\frac{A(X)dX}
                                 {V(X)(X-Y_\mu)}\right)=
                              \int\limits_{-1}^1 \frac{(-1)^q g(x^2)\,dx}
                              {(\mbox{i}x-Y_\mu)Z_D} \]  
\[\cdot\,\sum_{\nu=1}^q\left(\int_{X_{2\nu-1}}^{X^{(2\nu-1)}}\frac{A(X)dX}
{V(X)(X-\overline{Y}_\mu)}
                             +\int_{X_{2\nu}}^{X^{(2\nu)}}\frac{A(X)dX}
                             {V(X)(X-\overline{Y}_\mu)}   \right)=
                               \int\limits_{-1}^1 
                               \frac{(-1)^q g(x^2)\,dx}{(\mbox{i}x-
                               \overline{Y}_\mu)Z_D} \]  
Furthermore 
\begin{eqnarray}\nonumber\lefteqn{
\int_{X_{2\nu-1}}^{X^{(2\nu-1)}}\frac{A(X)dX}{V(X)(X-Y)}
+\int_{X_{2\nu}}^{X^{(2\nu)}}\frac{A(X)dX}{V(X)(X-Y)}=}\\ & &
 \nonumber -\int_{X_{2\nu}}^{X_{2\nu-1}}\frac{A(X)dX}
 {V(X)(X-Y)}+\int_{X^{(2\nu)}}^{X^{(2\nu-1)}}\frac{A(X)dX}{V(X)(X-Y)}
\end{eqnarray}
with $X^{(2\nu)}$ now lying in the other sheet of the Riemann surface.\\ 
In the limit $\varepsilon\to 0$, one obtains 
\[\lim_{\varepsilon\to 0}\,\int_{X_{2\nu}}^{X_{2\nu-1}}
\frac{A(X)dX}{V(X)(X-Y)}=\left\{
             \begin{array}{l} \pm\pi\mbox{i}\delta_{\mu\nu}
             /[\lambda_\mu(Y_\mu+\mbox{i}z)] \quad
             \mbox{for}\quad   Y=Y_\mu \\ [2mm]
             0\quad \mbox{for}\quad Y=\overline{Y}_\mu
             \end{array}\right.\]
with $\delta_{\mu\nu}$ being the usual Kronecker symbol 
and $\lambda_\mu$ as defined in (\ref{G1.7}).\\
The second term amounts to 
\begin{eqnarray}\nonumber\lefteqn{
\lim_{\varepsilon\to 0}\,\int_{X^{(2\nu)}}^{X^{(2\nu-1)}}
\frac{A(X)dX}{V(X)(X-Y)}=
\int_{X^{(2\nu)}}^{X^{(2\nu-1)}}\frac{dX}{(X-Y)\sqrt{(X+\mbox{i}z)
(X-\mbox{i}\bar{z})}}=  }\\[5mm] && \nonumber
\frac{1}{\lambda(Y)(Y+\mbox{i}z)}\ln\left(
        \frac{[\lambda(X^{(2\nu-1)})-\lambda(Y)]\,
        [\lambda(X^{(2\nu)})+\lambda(Y)]}
             {[\lambda(X^{(2\nu-1)})+\lambda(Y)]\,
             [\lambda(X^{(2\nu)})-\lambda(Y)]}\right),\nonumber \end{eqnarray}
where for evaluation of the second integral the substitution 
\[\lambda=\lambda(X)=\sqrt{\frac{X-\mbox{i}\bar{z}}
             {X+\mbox{i}z}}\] 
has been used.\\
Hence, the Jacobian inversion problem reads as follows in the 
limit $\varepsilon\to 0$:
\begin{eqnarray}\label{AG1a}
\cdot\,  \prod\limits_{\nu=1}^q\frac{[\lambda(X^{(2\nu-1)})-\lambda_\mu]
\,[\lambda(X^{(2\nu)})+\lambda_\mu]}
                     {[\lambda(X^{(2\nu-1)})+\lambda_\mu]\,
                     [\lambda(X^{(2\nu)})-\lambda_\mu]}&=&-\gamma_\mu  \\[5mm]
\label{AG1b}\cdot\,  \prod\limits_{\nu=1}^q\frac{[\lambda(X^{(2\nu-1)})-
\lambda_\mu^*]\,[\lambda(X^{(2\nu)})+\lambda_\mu^*]}                     
{[\lambda(X^{(2\nu-1)})+\lambda_\mu^*]\,[\lambda(X^{(2\nu)})-\lambda_\mu^*]}
&=&\overline{\gamma}_\mu
\end{eqnarray}                   
and in an analogous manner
\begin{eqnarray}\label{AG2}f=f_0\,\prod_{\nu=1}^q
\frac{[\lambda(X^{(2\nu-1)})+1]\,[\lambda(X^{(2\nu)})-1]}
                     {[\lambda(X^{(2\nu-1)})-1]\,
                     [\lambda(X^{(2\nu)})+1]} \end{eqnarray}
                     [with $\gamma_\mu,\lambda_\mu^*$ and $f_0$ as 
                     defined in (\ref{G1.7})].\\ 
Instead of evaluating the quantities $\lambda(X^{(\nu)}),\,(1\leq\nu\leq 2q),
$ the coefficients 
$b_\nu$ and $c_\nu\,(1\leq\nu\leq q)$ of the polynomial \\
\parbox{10cm}{
\begin{eqnarray}
\nonumber P(\lambda)&=&\prod_{\nu=1}^q[\lambda-
\lambda(X^{(2\nu-1)})]\,[\lambda+\lambda(X^{(2\nu)})]\\[5mm]&
=&\lambda^{2q}+\lambda\sum_{\nu=1}^q b_\nu\lambda^{2\nu-2}+
\sum_{\nu=1}^q c_\nu\lambda^{2\nu-2}\nonumber\end{eqnarray}}\hfill
\parbox{1cm}{\begin{eqnarray}\label{AG3}\end{eqnarray}}\\
are determined. 
Since 
\begin{eqnarray}\label{AG4}\frac{P(\lambda_\mu)}
{P(-\lambda_\mu)}=-\gamma_\mu,\quad
  \frac{P(\lambda_\mu^*)}{P(-\lambda_\mu^*)}=
  \overline{\gamma}_\mu,\quad
  f=f_0\frac{P(-1)}{P(1)},\end{eqnarray}
the following system of linear equations for the 
quantities $b_\nu,c_\nu,P(1),$ and $P(-1)$ emerges: \\
\parbox{10cm}{
\[\cdot\,\sum_{\nu=1}^q\left[b_\nu\alpha_\mu\lambda_\mu^{2\nu-1}
+c_\nu\lambda_\mu^{2\nu-2}\right]=-\lambda_\mu^{2q},\]
\[\cdot\,\sum_{\nu=1}^q\left[b_\nu\alpha_\mu^*(\lambda_\mu^*)^{2\nu-1}
  +c_\nu(\lambda_\mu^*)^{2\nu-2}\right]=-(\lambda_\mu^*)^{2q}\]
\[\cdot\,\sum_{\nu=1}^q(b_\nu-c_\nu)+P(-1)=1\]
\[\cdot\,\sum_{\nu=1}^q(b_\nu+c_\nu)-P(1)=-1,\] }\hfill
\parbox{1cm}{\begin{eqnarray}\label{AG5}\end{eqnarray}}\\
with $\alpha_\mu$ and $\alpha_\mu^*$ as defined in (\ref{G1.7}).\\
Finally, if the solution of this linear system for $P(\pm1)$ 
is expresseed by means of Cramer's rule, the desired 
form (\ref{G1.7}) of the B\"{a}cklund type is obtained.

\section{Invariance properties of the Ernst potential} 
For the proof of the properties (\ref{G2.1}-\ref{G2.2b}), the Ernst potential 
(\ref{G1.7}) is reformulated by
\begin{equation}f(\{Y_\nu\}_q\,;g)=f_0\frac{D(-1;\{Y_\nu\}_q;g)}{D(1;\{Y_\nu\}_q;g)}\label{IP}\end{equation}
with
\[ \begin{array}{l}  \cdot\, D(\lambda;\{Y_\nu\}_q;g)\\[5mm] \qquad\quad
                     =\left|\begin{array}{cccccccc} 
                           a_1&(a_1x_1)&\cdots&(a_1x_1^{q-1})
                           &1& x_1&\cdots&x_1^{q}\\[2mm]
                           a_2&(a_2x_2)&\cdots&(a_2x_2^{q-1})&1&
                            x_2&\cdots&x_2^{q}\\[2mm]
                           \vdots& \vdots&\ddots&\vdots&\vdots&
                           \vdots&\ddots&\vdots\\[2mm]
                           a_{2q+1}&(a_{2q+1}x_{2q+1})&\cdots&
                           (a_{2q+1}x_{2q+1}^{q-1})&1& x_{2q+1}
                           &\cdots&x_{2q+1}^{q}\\[2mm]
                    \end{array}\right|\\[1.5cm]
 \cdot\,a_1=\lambda\,,\quad a_{2\nu}=\alpha_\nu\lambda_\nu\,,
      \quad a_{2\nu+1}=\alpha_\nu^*\lambda_\nu^*\,,\\[2mm]

  \cdot\,x_1=\lambda^2\,,\quad x_{2\nu}=\lambda_\nu^2\,,
      \quad x_{2\nu+1}=(\lambda_\nu^*)^2.
 \end{array}\]
 The above expression for $D(\lambda;\{Y_\nu\}_q;g)$ is a Vandermonde-like determinant. 
 These determinants have been studied in detail by 
 Steudel, Meinel and Neugebauer \cite{SMN97}. By their {\em 
 reduction formula} [(8) of \cite{SMN97}], $D$ assumes the form:
\begin{eqnarray}\nonumber
 \lefteqn{D(\lambda;\{Y_\nu\}_q;g)}\\[3mm]&&\nonumber
  ={\cal V}_{q,q+1}(a_r;b_r|\,x_r)
  \qquad\mbox{[with $b_r=1$ for $r=1\ldots(2q+1)$]}\\[3mm]&&\nonumber
  =\sum_P\varepsilon_P\left(\,\prod_{j=1}^q a_{r(j)}\right)
  {\cal V}_q[x_{r(1)},\ldots,x_{r(q)}]\,
         {\cal V}_{q+1}[x_{r(q+1)},\ldots,x_{r(2q+1)}]\end{eqnarray}
where 
\begin{itemize}
\item[$\cdot$] the sum runs over all permutations $P=[r(1),\ldots,r(2q+1)]$ of $(1,2,\ldots,2q+1)$
with $r(k)<r(j)$ for $k<j<q$ as well as for $q\leq k<j$
\item[$\cdot$] $\varepsilon_P=\left\{\begin{array}{l} +1\quad\mbox{for $P$ even} \\-1 \quad\mbox{for $P$ odd}\end{array}\right.$
\item[$\cdot$] the Vandermonde determinants are given by\\ 
$\hspace*{1cm}{\cal V}_N[x_1,\ldots,x_N]=\prod\limits_{k>j}(x_k-x_j).$
\end{itemize}
In this formulation the following properties can be proved:
\begin{enumerate}
\item[(A)] If $x_{2q+1}=x_{2q}$ then 
\begin{eqnarray}
   \lefteqn{\nonumber D(\lambda;\{Y_\nu\}_q;g)}\\
   &&=(-1)^q(a_{2q}-a_{2q+1})\left[\,\prod_{j=1}^{2q-1}(x_{2q}-x_j)\right]
                              D(\lambda;\{Y_\nu\}_{q-1};-g)\nonumber
\end{eqnarray}
\item[(B)] If $x_{2q}=1+\kappa\epsilon+{\cal O}(\epsilon^2),\,x_{2q+1}=1-\kappa\epsilon+{\cal O}(\epsilon^2),$ 
           and $(a_{2q}a_{2q+1})=1+{\cal O}(\epsilon),$ then
      \begin{eqnarray}\lefteqn{D(\mp 1;\{Y_\nu\}_q;g)\nonumber}\\& &
      =\kappa\epsilon\left[\,\prod_{j=2}^{2q-1}(1-x_j)\right] 
      (a_{2q}+a_{2q+1}\pm2)D(\mp 1;\{Y_\nu\}_{q-1};g)
      +{\cal O}(\epsilon^2).\nonumber
      \end{eqnarray}
\end{enumerate}
With (A) the equalities (\ref{G2.1}) and (\ref{G2.2}) can be derived whilst 
(B) serves to confirm (\ref{G2.2a}) and (\ref{G2.2b}). 
In order to prove (A) consider the following groups of permutations separately:
\[\begin{array}{l}           
P_1: r(q-1)=2q,\,r(q)=2q+1\\[3mm]
P_2: r(2q)=2q,\,r(2q+1)=2q+1\\[3mm]
P_3: r(q)=2q,\,r(2q+1)=2q+1\\[3mm]
P_4: r(q)=2q+1,\,r(2q+1)=2q
\end{array}\]
For $x_{2q+1}=x_{2q}$, all terms belonging to $P_1$ and $P_2$ vanish while all 
terms belonging to $P_3$ and $P_4$ 
possess a common factor, $[a_{2q}\prod_{j=1}^{2q-1}(x_{2q}-x_j)]$ and 
$[a_{2q+1}\prod_{j=1}^{2q-1}(x_{2q}-x_j)]$, 
respectively. After reordering (from which the factor $(-1)^q$ results), 
(A) is easily obtained.\\
The proof for (B) works similarly. Now, eight groups of permutations have 
to be considered separately:
\[\begin{array}{l}           
P_{1a}: r(1)=1,\,r(q-1)=2q,\,r(q)=2q+1\\[3mm]
P_{1b}: r(q+1)=1,\,r(2q)=2q,\,r(2q+1)=2q+1\\[3mm]
P_{2a}: r(q)=2q,\,r(q+1)=1,\,r(2q+1)=2q+1\\[3mm]
P_{2b}: r(1)=1,\,r(q)=2q,\,r(2q+1)=2q+1\\[3mm]
P_{3a}: r(q)=2q+1,\,r(q+1)=1,\,r(2q+1)=2q\\[3mm]
P_{3b}: r(1)=1,\,r(q)=2q+1,\,r(2q+1)=2q\\[3mm]
P_{4a}: r(q-1)=2q,\,r(q)=2q+1,\,r(q+1)=1\\[3mm]
P_{4b}: r(1)=1,\,r(2q)=2q,\,r(2q+1)=2q+1\\[3mm]
\end{array}\]
All terms of permutations with a coinciding first index can be combined to 
give\footnote{Here the requirements $a_1=\mp 1,\,x_1=1$ are 
necessary. Additionally, for $P_{4a}$ and $P_{4b}$, the constraint 
$a_{2q}a_{2q+1}=1+{\cal O}(\epsilon)$ is needed.} :
\[\begin{array}{l}\{P_{1a},P_{1b}\}\Longrightarrow {\cal O}(\epsilon^3)\\[3mm]
                  \{P_{2a},P_{2b}\}\Longrightarrow a_{2q}F+{\cal O}(\epsilon^2)
                  \\[3mm]
                  \{P_{3a},P_{3b}\}\Longrightarrow a_{2q+1}F+{\cal O}
                  (\epsilon^2)\\[3mm]
                  \{P_{4a},P_{4b}\}\Longrightarrow \pm 2F+{\cal O}(\epsilon^2)
                  \end{array}\]
with \[F=(-1)^{q+1}\kappa\epsilon\prod_{j=2}^{2q-1}(1-x_j)D(\pm 1;
                  \{Y_\nu\}_{q-1};-g).\]
\section{Newtonian and ultrarelativistic limits} 
\subsection{The Newtonian limit}
In the limit of small functions $g$ and $\xi$, i.~e.~ 
\[g(x^2)=\varepsilon_g g_0(x^2)+{\cal O}(\varepsilon_g^2),\,\quad
  \xi(x^2)=\varepsilon_\xi \xi_0(x^2)+{\cal O}(\varepsilon_\xi^2),\]
the Ernst potential $f=f(\xi;g)$ as introduced in section 2 is given by\\
\parbox{10cm}{
\begin{eqnarray}\lefteqn{\nonumber f(\xi;g)=}\\&&\nonumber
1-\varepsilon_g\int\limits_{-1}^1\frac{g_0(x^2)dx}{Z_D}
                -\mbox{i}\,\varepsilon_g\varepsilon_\xi
                 \int\limits_{-1}^1\frac{(\mbox{i}x) g_0(x^2)\xi_0(x^2)dx}{Z_D} 
                 +{\cal O}(\varepsilon_g^2)+{\cal O}(\varepsilon_g
                 \varepsilon_\xi^2).\end{eqnarray}}
                 
\hfill
\parbox{1cm}{\begin{eqnarray}\label{AGNL1}\end{eqnarray}}\\                 
In this section, the above property will be proved and the functions
$g_0$ and $\xi_0$ will be derived as they result from the Newtonian 
expansion of the boundary conditions. 
\subsubsection{The Ernst potential for small functions $g$ and $\xi$}
Due to the assumption that the function $\Phi_g$ introduced in (\ref{Phi}) 
can be extended to 
form a continuous mapping defined on $\cal A$ (see sections 2 and 5), 
the representation of $\xi$ in terms of $\{Y_\nu\}_q$ can be chosen 
arbitrarily. 
Here, the following set $\{Y_\nu\}_q$ is used:
\begin{itemize}
\item[$\cdot$]  $q=4r$
\item[$\cdot$]  $\left\{\begin{array}{l} Y_{4\nu-3}=Z_\nu(1+\varepsilon_\xi 
z_\nu),\quad Y_{4\nu-2}=-\overline{Y}_{4\nu-3}\\[5mm]
                                  Y_{4\nu-1}=\overline{Z}_\nu
                                  (1-\varepsilon_\xi z_\nu),\quad
                                  Y_{4\nu}=-\overline{Y}_{4\nu-1}\end{array}
                                  \right\},\quad\begin{array}{l}
                                  \Re(Z_\nu)\neq 0,\quad z_\nu\in\mathbb{R}\\
                                  (\nu=1\ldots r)\end{array}$
\end{itemize}
Then, it follows from (\ref{G2.3}) that $\xi(x^2)=\varepsilon_\xi 
\xi_0(x^2)+{\cal O}(\varepsilon_\xi^2)$ with
\[\xi_0(x^2)=-4\mbox{i}\sum_{\nu=1}^{r}
       \frac{z_\nu(Z_\nu-\overline{Z}_\nu)(x^2-Z_\nu\overline{Z}_\nu)}
       {(x^2+Z_\nu^2)(x^2+\overline{Z}_\nu^{\;2})}.\]
To evaluate the Ernst potential in this limit, the formulation 
(\ref{AG1a}-\ref{AG5}) in appendix A is used and the 
following steps are performed:
\begin{enumerate}
\item At first, it turns out that in the limit $\varepsilon_\xi\to 0$ the 
coefficients $b_\nu$ of the polonomial 
(\ref{AG3}) vanish. This can be seen by considering the solution to linear 
system (\ref{AG5}).
\begin{eqnarray}\lefteqn{b_\nu=\frac{D_\nu}{D}\,:}\nonumber\\[3mm]&&\nonumber
\begin{array}{l}\cdot\quad D=
                \left|\begin{array}{ccccccc} 
                           a_2&\cdots&(a_2x_2^{q-1})&1& x_2&
                           \cdots&x_2^{q-1}\\[2mm]
                           \vdots& \ddots&\vdots&\vdots&\vdots&
                           \ddots&\vdots\\[2mm]
                           a_{2q+1}&\cdots&(a_{2q+1}
                           x_{2q+1}^{q-1})&1& x_{2q+1}&\cdots&x_{2q+1}^{q-1}
                           \\[2mm]
                    \end{array}\right|\\[1.5cm]
 \cdot\,\quad a_{2\eta}=\alpha_\eta\lambda_\eta\,,\quad a_{2\eta+1}=
 \alpha_\eta^*\lambda_\eta^*\,
        ,\quad x_{2\eta}=\lambda_\eta^2\,,\quad x_{2\eta+1}=
        (\lambda_\eta^*)^2\\[5mm]
 \cdot\quad \mbox{$D_\nu$ is derived from $D$ by replacing 
 the $\nu$-th column by the vector}\\[3mm]
 \qquad\qquad  \{-x_2^q,\ldots,-x_{2q+1}^q\}.
  \end{array}\end{eqnarray}
 For $1\leq\nu\leq q\,$, $\,D_\nu$  can be expanded in terms of Vandermonde 
 determinants 
 \[{\cal V}_{q+1}(x_{r(1)},\ldots,x_{r(q+1)}),\quad r(\eta)\in
 \{2,\ldots,2q+1\},\quad r(\eta)<r(\mu)\;\mbox{for}\; \eta<\mu\,. \]
In the limit $\varepsilon_\xi\to 0$, any set $\{x_{r(\eta)}\}_{q+1}$ 
contains at most $q$ different values, and therefore 
all $D_\nu$ vanish. On the other hand, $D$ remains finite (here only 
Vandermonde determinants ${\cal V}_q$ are involved), 
and hence all $b_\nu$ tend to zero.
\item Thus, with any zero $\tilde{\lambda}_\nu$ of the Polynomial (\ref{AG3}), 
$(-\tilde{\lambda}_\nu)$ also becomes a 
zero as $\varepsilon_\xi\to 0$. This set of zeros is ordered in the 
following way:
\[\{\lambda(X^{(1)}),-\lambda(X^{(2)}),\ldots,\lambda(X^{(2q-1)}),
-\lambda(X^{(2q)})\}=
  \{\tilde{\lambda}_1,-\tilde{\lambda}_1\ldots,\tilde{\lambda}_q,
  -\tilde{\lambda}_q\},\]
Suppose there is a $\lambda_\mu$ different from all zeros :
\[\lambda_\mu\neq\lambda(X^{(2\nu-1)})=\lambda(X^{(2\nu)})\quad\mbox{and}\quad
   \lambda_\mu\neq-\lambda(X^{(2\nu-1)})\quad\mbox{for all $\nu=1\ldots q$.}\]
Then, since $\gamma_\mu\neq-1$ for small $\,g$, 
(\ref{AG1a}) cannot be satisfied.
\item This gives rise to the following ansatz $(\nu=1\ldots q)$:
\[ \lambda^2(X^{(2\nu-1)})=\lambda^2_{\nu}+\varepsilon_\xi\kappa_{2\nu-1}
+{\cal O}(\varepsilon_\xi^2),\qquad
   \lambda^2(X^{(2\nu)})=\lambda^2_{\nu}+\varepsilon_\xi\kappa_{2\nu}+{\cal O}
   (\varepsilon_\xi^2),\]
 by which the system (\ref{AG1a}/\ref{AG1b}) can easily be solved to get the 
 set $\{\kappa_\nu\}_{2q}\,$. 
\item Finally, if $g(x^2)=\varepsilon_g g_0(x^2)+{\cal O}(\varepsilon_g^2)$ 
is considered, then (\ref{AGNL1}) follows
from (\ref{AG2}) by inserting the values obtained for 
$\{\lambda(X^{(\nu)})\}_{2q}\,$.
\end{enumerate}
\subsubsection{The functions $g_0$ and $\xi_0$ as resulting from the 
boundary conditions}
For any family of Ernst potentials $f=f(g_\varepsilon;\xi_\varepsilon)$ 
describing a sequence of differentially rotating 
disks of dust with the parameter $\varepsilon=M^2/J$ [$M$ and $J$ as 
defined in (\ref{AS2})], 
the following expansion is valid (see \cite{TKl}, pp. 83-89):
\[f=1+e_2(\rho,\zeta)\varepsilon^2+\mbox{i}b_3(\rho,\zeta)\varepsilon^3+
{\cal O}(\varepsilon^4).\]
By comparison with (\ref{AGNL1}) one gets
\[\cdot\,\varepsilon_g=\varepsilon^2,\quad\varepsilon_\xi=\varepsilon,\]
\[\cdot\,e_2(\rho,\zeta)=-\int\limits_{-1}^1\frac{g_0(x^2)dx}{Z_D},\quad 
               b_3(\rho,\zeta)=-\int\limits_{-1}^1\frac{(\mbox{i}x) g_0(x^2)
               \xi_0(x^2)dx}{Z_D}.\]
If the boundary conditions, 
\begin{eqnarray}\cdot\,\sigma_p(\rho)&=&\sigma_0\psi_2[(\rho/\rho_0)^2]
\sqrt{1-(\rho/\rho_0)^2}\,
  \varepsilon^2+{\cal O}(\varepsilon^4)\quad\mbox{(with $\psi_2(0)=1$)}
  \qquad\mbox{or}\nonumber\\
  \cdot\,\Omega(\rho)&=&\Omega_0\Omega_1[(\rho/\rho_0)^2]\,\varepsilon+
  {\cal O}(\varepsilon^3)
  \quad\mbox{(with $\Omega_1(0)=1$)},\nonumber\end{eqnarray}
are given, then it follows from equations (\ref{G1.3}-\ref{G1.6}) that
\begin{eqnarray} \cdot\,(e_2)_{,\zeta}&=&4\pi\sigma_0\psi_2
\sqrt{1-(\rho/\rho_0)^2}\quad\mbox{or}\quad
     (e_2)_{,\rho}=2\Omega_0^2\Omega_1^2\,\rho\quad\mbox{and}\nonumber\\
     \cdot\, (b_3)_{,\rho}&=&2\rho\,\Omega_0\Omega_1\,(e_2)_{,\zeta}.
     \quad\nonumber\end{eqnarray}
By expressing $e_2$ and $b_3$ in terms of $g_0$ and $\xi_0$ in these 
equations, one gets Abelian 
integral equations for $\xi_0$ and $g_0$. Their solutions read as follows:
\begin{eqnarray}
 g_0(x^2)&=&-4\sigma_0(1-x^2)\int_0^{\pi/2}(\sin^2\phi)\,\psi_2(
 \cos^2\phi+x^2\sin^2\phi)\,d\phi\nonumber\\
g_0(x^2)\xi_0(x^2)&=&8\sigma_0\Omega_0(1-x^2)\int_0^{\pi/2}(\sin^2\phi)
  \,\tilde{\Omega}_1(\cos^2\phi+x^2\sin^2\phi)\,d\phi\nonumber\\& &
\hfill\mbox{[with $\tilde{\Omega}_1(x^2)=\Omega_1(x^2)\psi_2(x^2)].$}
\nonumber\end{eqnarray}
Note that only {\em one} of the functions $\psi_2$ and $\Omega_1$ can be 
prescribed since both represent
different boundary conditions of the same Newtonian potential $e_2$. 
Likewise, the constants $\sigma_0$ and $\Omega_0^2$ 
depend on each other. Moreover, these constants in terms of $\psi_2$ 
and $\Omega_1$ are prescribed by the equation 
$\varepsilon=M^2/J$.
\subsection{The ultrarelativistic limit} 
It is difficult to relate the functions $g$ and $\xi$ of an Ernst potential 
$f=f(g;\xi)$ to its physical properties like 
$M$ and $J$. Nevertheless, if a sequence $f(g_\varepsilon;\xi_\varepsilon)$ 
can be extended to arbitrary values 
$\varepsilon<1$, then, in the limit $\varepsilon\to 1$, the universal solution 
of an extreme Kerr black hole is 
reached. It is illustrated how this limit results from the form (\ref{G1.7}) 
of the Ernst potential.\\[5mm]
If the limit $\rho_0\to 0$ is considered for finite values of $r=\sqrt{\rho^2
+\zeta^2}$, then by using the 
formulation (\ref{IP}) one gets (with $\zeta=r\cos\theta$):
\[f=\left(1-\frac{\rho_0}{r}\int\limits_{-1}^1(-1)^qg(x^2)dx+{\cal O}(\rho_0^2)
\right)\!\!\left[
\frac{E_1r+\rho_0[E_3\cos\theta-(-1)^qE_2]}
{E_1r+\rho_0[E_3\cos\theta+(-1)^qE_2]}+{\cal O}(\rho_0^2)\right].\]
The $E_j$ do not depend on $\rho$ and $\zeta$ but on $g$ and $\xi$. 
In particular:
\[ \cdot\, E_1=\left|\begin{array}{cccccccc} 
                           b_1&(b_1Z_1)&\cdots&(b_1Z_1^{q-1})&1& Z_1&
                           \cdots&Z_1^{q-1}\\[2mm]
                           \vdots& \vdots&\ddots&\vdots&\vdots&\vdots&
                           \ddots&\vdots\\[2mm]
                           b_{2q}&(b_{2q}Z_{2q})&\cdots&(b_{2q}Z_{2q}^{q-1})
                           &1& Z_{2q}&\cdots&Z_{2q}^{q-1}\\[2mm]
                    \end{array}\right| \]
\[ \cdot\, E_2=\left|\begin{array}{cccccccc} 
                           b_1&(b_1Z_1)&\cdots&(b_1Z_1^{q-2})&1& Z_1&
                           \cdots&Z_1^{q}\\[2mm]
                           \vdots& \vdots&\ddots&\vdots&\vdots&\vdots&
                           \ddots&\vdots\\[2mm]
                           b_{2q}&(b_{2q}Z_{2q})&\cdots&(b_{2q}Z_{2q}^{q-2})
                           &1& Z_{2q}&\cdots&Z_{2q}^{q}\\[2mm]
                    \end{array}\right| \]
\[\cdot\, b_{2\nu-1}=-\tanh\left[\frac{1}{2}\int\limits_{-1}^1\frac{(-1)^q
g(x^2)dx}{\mbox{i}x-Y_\nu}\right]
 \,,\quad b_{2\nu}\overline{b}_{2\nu-1}=1\]
\[\cdot\,Z_{2\nu-1}=Y_\nu\,,\quad Z_{2\nu}=\overline{Y}_\nu.\]
 Clearly, if $E_1\neq 0$ then $\lim_{\rho_0\to0}f=1$. The Ernst potential 
 passes to an ultrarelativistic limit
 if $E_1$ and $\rho_0$ tend simultaneously to zero such 
 that\footnote{It can 
 be shown that $E_1^2\in\mathbb{R}$. Hence, the ultrarelativistic limit 
 for the family $f(g_\varepsilon;\xi_\varepsilon)$
 is performed when some function $E_a=E_a(g_\varepsilon;\xi_\varepsilon)=
 E_b(\varepsilon)E_1^2(\varepsilon)$, which is 
 independent of the representation $\{Y_\nu\}_q$, vanishes.}
 \[\Omega_U=\lim_{\rho_0\to0}\frac{(-1)^qE_1}{\,2E_2\rho_0}\]
 exists. Then one gets
 \[ f=\frac{2\Omega_Ur+E_4\cos\theta-1}{2\Omega_Ur+E_4\cos\theta+1}\,.\]
The only Ernst potential of this form which is asymptotically flat and 
regular for $r>0$ is the extreme Kerr solution. 
The constant $\Omega_U$ is then real and describes the `angular velocity 
of the horizon'. Moreover, 
$J=1/(4\Omega_U^2)=M^2$, and hence $\varepsilon=1$.
\end{appendix}
\section*{ACKNOWLEDGEMENTS}
The author would like to thank A.~Kleinw\"{a}chter, R. Meinel, and 
G.~Neugebauer for many valuable discussions. \\The support from the DFG 
is gratefully acknowledged.


\begin{thebibliography}{99}
\bibitem{AM00} Ansorg, M., and Meinel, R. (2000). gr-qc/9910045, to appear 
in {\it Gen. Rel. Grav.}
\bibitem{MN96}  Meinel, R., and Neugebauer, G. (1996). {\it Phys. Lett. A} 
{\bf 210}, 160.
\bibitem{Kor89} Korotkin, D.~A. (1989). {\it Theor. Math. Phys.} 
{\bf 77}, 1018.
\bibitem{Kor93} Korotkin, D.~A. (1993). {\it  Class. Quantum Grav.}  
{\bf 10}, 2587.
\bibitem{Kor97} Korotkin, D.~A. (1997). {\it  Phys. Lett. A} {\bf 229}, 195.
\bibitem{MN97} Meinel, R., and Neugebauer, G. (1997). {\it Phys. Lett. A} 
{\bf 229}, 200.
\bibitem{NM93} Neugebauer, G., and Meinel, R. (1993). 
{\it Astrophys. J.} {\bf 414}, L97.
\bibitem{NM95} Neugebauer, G., and Meinel, R. (1995). 
{\it Phys. Rev. Lett.} {\bf 75}, 3046.
\bibitem{NKM96} Neugebauer, G., Kleinw\"{a}chter, A., 
and Meinel, R. (1996). 
               {\it Helv. Phys. Acta} {\bf 69}, 472.

\bibitem{Mais78} Maison, D. (1978). {\it Phys. Rev. Lett.} 
{\bf 41}, 521.
\bibitem{Bel78} Belinski, V.~A., and Zakharov, V.~E. (1978). 
{\it Zh. Eksper. Teoret. Fiz.} 
   {\bf 75}, 195.
\bibitem{Har78} Harrison, B.~K. (1978). {\it Phys. Rev. Lett.} {\bf 41}, 119.
\bibitem{Herlt78} Herlt, E. (1978). {\it Gen. Rel. Grav.} {\bf 9}, 711.
\bibitem{Hoense79} Hoenselaers, C., Kinnersley, W., 
                   and Xanthopoulos, B.~C. (1979). {\it Phys. Rev. Lett.} 
                   {\bf 42}, 481.
\bibitem{N79} Neugebauer, G. (1979). {\it J. Phys. A} {\bf 12}, L67.
\bibitem{N80a} Neugebauer, G. (1980). {\it J. Phys. A} {\bf 13}, L19.
\bibitem{N80b} Neugebauer, G. (1980). {\it J. Phys. A} {\bf 13}, 1737.
\bibitem{Hausera} Hauser, I. and Ernst, F.~J. (1979). {\it Phys. Rev D} 
{\bf 20}, 362.
\bibitem{Hauserb} Hauser, I. and Ernst, F.~J. (1979). {\it Phys. Rev D} 
{\bf 20}, 1783.
\bibitem{Hauserc} Hauser, I. and Ernst, F.~J. (1980). {\it J. Math. Phys.} 
{\bf 21}, 1418.

\bibitem{N96} Neugebauer, G. (1996). In {\it General Relativity}, 
eds. G.~S.~Hall and J.~R.~Pulham,
Proceedings of the 46. Scottish Universities Summer School in Physics, 
Aberdeen, July 1995, pp. 61-81.
\bibitem{Ernst} Ernst, F.~J. (1968). {\it Phys. Rev.} 
{\bf 167}, 1175.
\bibitem{Ernsta} Kramer, D., and Neugebauer, G. (1968). {\it Commun. Math. 
Phys.} {\bf 7}, 173.

\bibitem{TKl} Kleinw\"{a}chter, A. (1995). ''Untersuchungen zu 
rotierenden Scheiben in der 
                    Allgemeinen Relativit\"{a}tstheorie.'' Ph.D. Dissertation, 
                    Friedrich-Schiller-Universit\"{a}t Jena.
\bibitem{Kerr63} Kerr, R. (1963). {\it Phys. Rev. Lett.} {\bf 11}, 237.
\bibitem{BW} Bardeen, J.~M., and Wagoner, R.~V. (1971). {\it  Astrophys. J.} 
{\bf 167}, 359.
\bibitem{Brandt} Brandt, J.~C.~ (1960). {\it Astrophys. J.} {\bf 131},293.
\bibitem{N00} Neugebauer, G.~(2000). {\it Ann. Phys. (Leipzig)} {\bf 9}, 3-5, 
342.
\bibitem{Aleks99} Alekseev, G.~A.~(2000). {\it Ann. Phys. (Leipzig)} {\bf 9}, 
Spec. Issue, SI-17 and references therein.

\bibitem{SMN97} Steudel, H., Meinel, R., Neugebauer, G. (1997). 
{\it  J. Math. Phys.} {\bf 38} (9), 4692.

\end{thebibliography}
\end{document}